\documentclass[sigconf,screen]{acmart}

\AtBeginDocument{%
  \providecommand\BibTeX{{%
    \normalfont B\kern-0.5em{\scshape i\kern-0.25em b}\kern-0.8em\TeX}}}

\copyrightyear{2024}
\acmYear{2024}
\setcopyright{rightsretained}
\acmConference[ARES 2024]{The 19th International Conference on Availability, Reliability and Security}{July 30 -- August 2, 2024}{Vienna, Austria}
\acmBooktitle{The 19th International Conference on Availability, Reliability and Security (ARES 2024), July 30 -- August 2, 2024, Vienna, Austria}\acmDOI{10.1145/3664476.3670433}
\acmISBN{979-8-4007-1718-5/24/07}

\usepackage{pifont}

\usepackage[frozencache]{minted}
\setminted{
    linenos,
    numbersep=5pt,
    xleftmargin=4ex,
    python3=true,
    frame=lines,
    framesep=2mm,
    breaklines=true,
}

\usepackage{url}
\usepackage{hyperref}
\usepackage[hyphenbreaks]{breakurl}

\newcommand{\cmark}{$\bullet$}%

\newcommand{\xmark}{$\circ$}%

\begin{document}

\title[If It Looks Like a Rootkit and Deceives Like a Rootkit]{If It Looks Like a Rootkit and Deceives Like a Rootkit: A~Critical~Examination of Kernel-Level Anti-Cheat Systems}

\author{Christoph Dorner}
\orcid{0009-0005-6085-3210}
\email{christoph.dorner@fhstp.ac.at}
\affiliation{%
    \institution{St.\ Pölten University of Applied Sciences}
    \city{St.\ Pölten}
    \country{Austria}
}

\author{Lukas Daniel Klausner}
\orcid{0000-0003-3650-9733}
\email{mail@l17r.eu}
\affiliation{%
    \institution{St.\ Pölten University of Applied Sciences}
    \city{St.\ Pölten}  
    \country{Austria}
}

\renewcommand{\shortauthors}{Dorner and Klausner}

\begin{abstract}
    Addressing a critical aspect of cybersecurity in online gaming, this paper systematically evaluates the extent to which kernel-level anti-cheat systems mirror the properties of rootkits, highlighting the importance of distinguishing between protective and potentially invasive software. After establishing a definition for rootkits (making distinctions between rootkits and simple kernel-level applications) and defining metrics to evaluate such software, we introduce four widespread kernel-level anti-cheat solutions. We lay out the inner workings of these types of software, assess them according to our previously established definitions, and discuss ethical considerations and the possible privacy infringements introduced by such programs. Our analysis shows two of the four anti-cheat solutions exhibiting rootkit-like behaviour, threatening the privacy and the integrity of the system. This paper thus provides crucial insights for researchers and developers in the field of gaming security and software engineering, highlighting the need for informed development practices that carefully consider the intersection of effective anti-cheat mechanisms and user privacy.
\end{abstract}




\maketitle

\section{Introduction}
Online video gaming, particularly competitive gaming and esports, consistently has to confront the issue of cheating. This challenge has intensified with technological advancements in gaming, prompting a response from game developers in the form of sophisticated anti-cheat systems. Notably, kernel-level anti-cheat systems have emerged as a solution to this issue; such systems operate at a deep system level, a characteristic they share with rootkits. Rootkits, known for stealth and deep system integration, typically serve unauthorised purposes, contrasting with the intended protective role of anti-cheat systems.

This paper examines the similarities and differences between kernel-level anti-cheat systems and rootkits. It begins by outlining the progression of cheating methods in online gaming and the corresponding evolution of anti-cheat technologies, focussing on those operating at the kernel level. The discussion then shifts to rootkits, defining their key characteristics for comparison. The primary analysis involves applying a set of metrics, developed for classifying rootkits, to various kernel-level anti-cheat systems used in current popular video games. This methodology is employed to assess whether these anti-cheat systems exhibit rootkit-like properties, potentially raising concerns about user privacy and system integrity.

The aim of this paper is to investigate the ethical and technical implications of kernel-level anti-cheat systems in online gaming. It seeks to address the balance between effective cheating prevention and safeguarding user privacy, a critical issue in the landscape of digital games. Our contributions are as follows:
\begin{itemize}
    \item We propose a set of criteria to characterise the rootkit-ness of a piece of software.
    \item We analyse the most prevalent kernel-level anti-cheat systems currently in use in online games and evaluate them based on these criteria, finding two of the four to be rootkit-like.
    \item Finally, we discuss how to balance security and privacy concerns when designing and employing kernel-level anti-cheat software in digital games.
\end{itemize}

The rest of this paper is structured as follows: First, we briefly discuss the limited existing literature on the topic. We then give a short background on rootkit and lay out the metrics for rootkit-like behaviour we have synthesised from the field of rootkit research. Thereafter, we discuss, analyse and evaluate the four most prevalent kernel-level anti-cheat systems used in online gaming (BattlEye, Easy Anti-Cheat, FACEIT Anti-Cheat and Vanguard) against these metrics, finding FACEIT Anti-Cheat and Vanguard to exhibit clear rootkit-like behaviour. Finally, we summarise our findings and discuss their implications, in particular with regard to taking user privacy into account when designing anti-cheat software.

This article thus contributes to the existing body of knowledge by providing insight into the methods utilised by kernel-level anti-cheat systems, classifying them as rootkits or non-rootkits based on their intrusiveness. The findings serve as a starting point for further investigations and discussion on the delicate balance between cheat detection and the preservation of user privacy and system integrity within the gaming community.

\section{Related Work}
Although there are numerous rootkits and kernel-level anti-cheat solutions, there has thus far been very little research focussing on examining and contrasting these two types of software. While several papers have been published discussing rootkits (which we discuss in more detail in \autoref{sec:methods}), their detection methods and typical behaviours, there is a noticeable lack of research on anti-cheat systems, with the limited existing literature thereon (much of it in the form of bachelor's and master's theses) often lacking in-depth analysis and comprehensive evaluation.

Rendenbach~ \cite{bakk:Rendenbach} defines a basic framework for game developers to discern which kinds of anti-cheat system best meet the demands of the game. This framework includes a number of metrics, such as the ethical stance on cheating of the developers, the available budget, privacy concerns and other relevant factors. By considering these metrics, game developers can make informed decisions on implementing an anti-cheat system that aligns with the specific requirements of the game and their ethical considerations. Following this framework, the author provides an overview of possible options that encompasses existing anti-cheat systems as well as other technical and psychological options. This comprehensive overview explores the various measures available and their effectiveness in addressing cheating in games. The discussion concludes with the proposal of a “trustworthiness” factor, which combines the presented measures to assess the overall reliability and efficacy of an anti-cheat system.

Lehtonen~\cite{masters:lethonen} provides an overview of the historical development of anti-cheat systems in video games. The study categorises and examines different techniques employed by anti-cheat systems. Additionally, the research evaluates and describes the effectiveness of combined approaches utilised by these systems. By presenting this comprehensive analysis, Lehtonen contributes to understanding anti-cheat systems in video games, their evolution and the evaluation of combined approaches for enhancing gameplay integrity and fair competition.

Silva~\cite{masters:silva} elaborates on anti-cheat methods solely residing on the server to increase users' privacy. Due to many games having the ability to replay matches, these demos can be analysed in retrospect, detecting any occurrences of cheating in the match.

Maario et al.~\cite{paper:RedefiningKernelLevelACs} propose the need for less intrusive anti-cheat systems and introduce an approach for such a system. In the paper, several types of cheats and anti-cheat measures are elaborated. Mirroring Silva, the conclusion of this paper also favours a server-sided anti-cheat mechanism utilising recorded games and evaluating them based on machine-learning approaches.

Beegle~\cite{article:rootkits2} presents an overview of the evolution of rootkits over time and gives a detailed classification of virtualised rootkits, categorising them based on specific characteristics and behaviours. Additionally, the study addresses the resulting threats to information security, covering prevention strategies, detection methods and reliable removal techniques. The study thus provides a comprehensive understanding of rootkits, their various types and the associated challenges in ensuring robust information security.

Jiang's work~\cite{phd:rootkits1} introduces a virtualised honeypot farm to attract and capture rootkit attacks to obtain samples from real-world incidents. This honeypot farm collects malware samples from the internet for subsequent analysis. The study also discusses how to design a highly efficient and accurate malware logger system. This logger system is aimed at effectively recording and analysing malware activities to enhance the understanding and response to such threats. By combining the virtualised honeypot farm and the malware logger, Jiang's research contributes to the field of cybersecurity by providing valuable insights into rootkit attacks and subsequent prevention methods and facilitating the analysis of captured malware samples. However, not all rootkits would execute on the virtualised platform.


\section{Methodology}\label{sec:methods}
In response to growing concerns within the online gaming community about the similarities between kernel-level anti-cheat systems and rootkits, this study employs a structured analytical approach. We focus on comparing prominent kernel-level anti-cheat systems against defined metrics associated with rootkits. The list of metrics used in this study is derived from an analysis of different studies concerning the behaviour of common rootkits, while the evaluation of the different anti-cheat solutions is based on empirical analyses of both security researchers and cheat developers. This approach enables us to portray the systems from different perspectives. An anti-cheat solution was only flagged as exhibiting a certain characteristic if clear evidence of rootkit-like behaviour was present. We selected the four kernel-level anti-cheat solutions analysed for analysis which are currently most prevalent and widely used in the online gaming community.

\subsection{Rootkits}\label{sec:rootkits}
Rootkits are a type of malware that provide cybercriminals with unauthorised access to a computer system while remaining hidden. They are classifiable into several types: firmware rootkits, kernel-mode rootkits, virtualised rootkits and user-mode rootkits (although these ones are not relevant to this study). 

\textbf{Firmware rootkits}~\cite{Li2011AnOO} target the software that runs particular hardware components, such as the BIOS, and are stored on the software that runs during the boot process before the operating system (OS) starts. They are stealthy as they can persist through OS reinstallation because the firmware on the hardware itself is infected.

\textbf{Kernel-mode rootkits}~\cite{Joy2011RootkitDM} modify the OS by adding or changing existing code. They often exploit the fact that OSes allow device drivers or loadable modules to execute with the same privileges as the OS kernel. To this end, they are often packaged as device drivers or modules to avoid detection.

\textbf{Virtualised rootkits}~\cite[p.\ 9]{thesis:rootkits2}\label{subsec:VirtualisedRootkits} boot up before the operating system and operate as malware that executes as a hypervisor controlling one or multiple virtual machines (VMs). This shimming between hardware and OS enables the rootkit to conduct its malicious activities with less chance of being detected since all VMs linked to the hypervisor appear to function normally. 



\subsection{Rootkit Metrics}\label{sec:rootkitMetrics}
After this brief description of different rootkit types, we now define a set of metrics to classify software as rootkit-like. We have synthesised a set of seven properties commonly attributed to rootkits in our classification scheme.

\textbf{Evasion:} Evasion and deception of intrusion prevention systems or standard antivirus scanners is a known feature of rootkits. Rootkits often accomplish this goal by hiding malicious processes or tampering with applications on the infected system, patching the binaries so they fail to show files created by the rootkit, and often employing hooking tactics~\cite{Leian:1}. If the rootkit starts before the OS has finished booting (as is the case in firmware rootkits), it can evade all detection mechanisms of the OS altogether due to it having control of all drivers and applications starting after itself~\cite{Eresheim2017}.
In addition, rootkits often attempt to disguise themselves as a different process using process hollowing~\cite[p.\ 28]{thesis:RootkitStealthTactics}. In this strategy, the memory of harmless processes is deallocated and overwritten by the rootkit for its usage while still operating under the name of the harmless process.
Some rootkits can also hide the generated network traffic amongst other features by manipulating the \mintinline{bash}|ioctl| syscall, typically used to read or change the driver parameters for attached devices, including the network card \cite[p.\ 244\,ff.]{blunden2013rootkit}.

\textbf{Virtualisation:} Rootkits often use virtualisation for evasion tactics and concealment of their code. 
Virtualisation is used to protect the image itself by using proprietary, custom architecture emulations. Often tools like \textit{VMProtect}\footnote{\ \href{https://vmpsoft.com/}{https://vmpsoft.com/}} are used to protect software images, making it difficult to reverse-engineer such software. Despite the use of these measures, it is possible to (partially) dump such modules from loaded memory areas or with the use of third-party tools to de-virtualise the images, enabling reverse-engineering them nonetheless~\cite{rolles2009unpacking}. Despite sharing the same terminology, this kind of virtualisation is different from what is used in virtualised rootkits.
 

\textbf{Time of execution:} Firmware rootkits are designed to start execution concurrently with or before the OS of the infected system, depending on the methods utilised by the rootkit. Booting this early leads to the advantage of hiding from user-level monitoring tools and the OS's security mechanisms. Additionally, this leads to the malicious software being able to eavesdrop on every action happening above it by hooking various methods, using syscalls, and manipulating loaded libraries. With these capabilities, the rootkit can monitor any function executed in the user and kernel space, including alterations \cite[p.\ 52\,f.]{matrosov2019rootkits}.

\textbf{Remote access and controllability:} Most rootkits implement methods for the threat actor to control and access the infected system remotely and often include infected targets in botnets. 
Infected machines in such networks become bots. Botnets are rentable to run large distributed denial-of-service attacks on various targets. Bots connect to a centralised command-and-control (C2) server infrastructure, which sends commands to the infected systems that the bot executes. 
Furthermore, an attacker can install additional malware onto the infected system using the remote access feature and disable or remove security features, further compromising the infected systems~\cite[p.\ 319]{matrosov2019rootkits}. This feature is often used to stage supplementary attacks on adjacent systems to perform lateral movement in the network. 

\textbf{Information exfiltration:} Rootkits often include functionalities to exfiltrate data from infected systems, especially targeting corporations or governments to steal classified or high-value information. Depending on the information stolen, it can be sold (e.\,g.\ on the dark web) or used for the attackers' gain. These techniques can also be used to obtain basic information about the system or the attached network to plan for further attacks and select suitable malware specifically targeting the infected systems' weaknesses~\cite[p.\ 80\,f.]{major2015taxonomic}.

\textbf{Network manipulation:} Given that the network card of a system requires drivers to operate, a rootkit can attach to these drivers by hooking functions or the syscall table directly to send and receive packets on the network. This technique enables the rootkit to stay undetected for user-mode applications and spoof the outgoing network connections, hiding the ones originating from the rootkit. Furthermore, monitoring and alteration of established connections are possible, often used to attack systems in the adjacent network or connect to a C2 server. This is especially useful for passively eavesdropping on the network.

\textbf{Removability:} Kernel-level rootkits can often be hard to remove, requiring specifically designed tools. It is even possible that the system needs to be reinstalled from scratch to remove all traces of the rootkit, which is the generally accepted best-practice approach to treat infected systems~\cite{article:rootkits2}. In some cases, replacing some or all system hardware components is necessary to remove a firmware rootkit entirely.

\subsection{Anti-Cheat Solutions}
The selection of kernel-level anti-cheat systems was based on those used in online games with the highest player counts (as recorded on \href{https://steamcharts.com}{steamcharts.com} and \href{https://tracker.gg/}{tracker.gg} on 21 January 2023). Additionally, the level of intrusiveness was considered, determined by the extent of controversy surrounding their functionality within the gaming community, as reflected in discussions on online forums\footnote{\ \href{https://www.reddit.com/r/pcmasterrace/comments/19acyup/is_anyone_concerned_with_more_and_more_gaming/}{https://www.reddit.com/r/pcmasterrace/comments/19acyup}}. This resulted in the anti-cheat systems of the games \textit{Counter-Strike: Global Offensive}, \textit{Fortnite}, \textit{Tom Clancy's Rainbow Six Siege} and \textit{Valorant} being chosen. All four systems are partially based in kernel mode and use similar methods. Due to the intrusiveness of modern kernel-level anti-cheat solutions, the boundaries between them and rootkits are often slim, causing a rogue kernel-level anti-cheat system to have the potential to devastate the systems it is installed on. We have chosen a simple point system for our evaluation, with a single point added to the rootkit-ness score of an anti-cheat system for each matching criterion. Summing these points after the analysis, we arrive at a score from zero to seven; if any given anti-cheat system is assessed with four or more points, we consider it to be a rootkit-like system.

\paragraph{BattlEye}
This German anti-cheat solution is well established in the gaming industry and used by various games from different genres, including \textit{Rainbow Six Siege}, \textit{PUBG: Battlegrounds} and others. The self-proclaimed “gold standard” of anti-cheat solutions must be installed on the server hosting the game and on all clients playing it, giving it an in-depth overview and control of all in-game events. 
The protection mechanisms of BattlEye are circumventable relatively easily. Thus, many sources reveal this anti-cheat solution's code and inner workings. BattlEye employs an account-based recognition system, which can issue global bans to rid a game of cheaters using multiple accounts.\footnote{\ \href{https://www.battleye.com/about/}{https://www.battleye.com/about/}} 

\paragraph{Easy Anti-Cheat}
Easy Anti-Cheat (EAC) is more sophisticated than BattlEye, which makes it harder to reverse-engineer and discover the mechanisms of this solution. This anti-cheat is owned and developed by Epic Games and built into games like \textit{Apex Legends}, \textit{Fortnite} or \textit{Rust}, amongst many others. EAC is free to use and easy to implement in any game for developers, which explains the popularity gained in recent years. Using the hardware ID (HWID) of the system, EAC can ban cheaters globally, increasing the difficulty for players with an unfair advantage to create new accounts and continue their cheating spree. 

\paragraph{FACEIT Anti-Cheat}
FACEIT Anti-Cheat is a third-party platform for playing games with a custom ranking system, often considered superior to those built into the games. \textit{Counter-Strike: Global Offensive} is one of the provided games, serving up to 22~million players. It is considered a rather sophisticated anti-cheat system by other reverse engineers in the field \footnote{\ \href{https://guidedhacking.com/threads/anticheat-faceit-bypass.16113/}{https://guidedhacking.com/threads/anticheat-faceit-bypass.16113/}} and bypassing the security measures it implements is an exceedingly challenging task.\footnote{\ \href{https://www.faceit.com/en}{https://www.faceit.com/en}} 

\paragraph{Vanguard}
The anti-cheat software Vanguard, developed by Riot Games, is a more intrusive solution. To ensure the successful launch of the protected games, the system must start in a secured state, necessitating the kernel driver to initialise during the boot process. Without this initialisation, the protected games fail to launch. At present, Vanguard is exclusively utilised for the game \textit{Valorant}, but there are plans to extend its usage to include other games developed by the studio in the future~\cite{web:VanguardOverview}.\\

\section{Results}\label{sec:results}
We now turn to analysing anti-cheat solutions for their rootkit-ness based on the metrics defined in \autoref{sec:rootkitMetrics}. This involves a detailed examination of four anti-cheat solutions, resulting in their classification based on these metrics.

\subsection{BattlEye}\label{sec:BattlEye}
BattlEye consists of four components~\cite{web:BattlEyeSecretClub}, working together to detect and ban cheaters: 
\begin{description}
    \item[BEService.] BEService is part of the Infrastructure that enables communication to the BattlEye server from the user- and kernel-mode parts installed on the client.
    \item[BEDaisy.] BEDaisy is the kernel-mode driver, allowing the anti-cheat system to enumerate the memory, scan for vulnerable libraries or insert other modules streamed from the server.
    \item[BEClient.] BEClient is the user-mode part of the anti-cheat system, injected into the game upon startup, and issues most calls to the driver to execute the various scans.
    \item[BEServer.] BEServer must be running on the game server the users connect to, where it can monitor user behaviour, receive all notifications from the BattlEye clients (through the BEClient), and issue bans for either the specific server or even globally.
\end{description} 
These four parts constitute the base of the BattlEye anti-cheat system. Game developers can use the API for the modules in the game code, embedding it deep into the game. This yields monitoring results for in-game player behaviour, making the detection of blatant cheats easy. It interfaces with the kernel driver BEDaisy, which can observe low-level memory areas, check for existing function hooks, etc. To evade tampering from cheats operating at the user level, the anti-cheat system performs scans within the kernel. The client gets the modules streamed from the BEServer, inserting them directly into the memory where they are allocated and executed. 
Reporting players for any violation is implemented in the BEService part of BattlEye via the function \mintinline{c}|battleye::report()|, which encapsulates information about the player and the infringement conducted. When a player gets reported, it does not mean a ban is immediately issued. However, more scanning modules will be streamed to the client, gathering additional information about the system. BattlEye can hence be characterised as a relatively lenient anti-cheat system. It primarily issues outright bans only when players are caught cheating by deliberately evading scans and protection measures. The usage of e.\,g.\ macros will, at most, lead to being kicked out of the current game or the game not starting.\footnote{\ \href{https://www.battleye.com/support/faq/}{https://www.battleye.com/support/faq/}}  

\subsubsection{BattlEye Anti-Cheat Measures:}\label{subsec:BattlEyeBasics}
The components of BattlEye use the named pipe \mintinline{text}{\.\namedpipe\Battleye} for inter-process communication, encrypting the traffic sent with a \mintinline{bash}|XOR| encryption using weak keys, which makes it vulnerable for known-plaintext attacks. This is the primary communication method between the components BEDaisy, BEService and BEClient. 

\paragraph{Memory Scans}
To detect cheats and anomalies in user behaviour, the kernel driver scans the whole memory area of the protected game and the anti-cheat system, allowing it to check for known cheats loaded. 
Additionally, all external memory pages are examined to determine whether they have the \mintinline{text}|X| (execute) bit set. This implies that any process with sufficient privileges can execute the code residing in these pages. These pages can be utilised to inject cheats or a loader into the memory, which subsequently can be used to exploit vulnerabilities in the game or circumvent the anti-cheat system \cite{web:BattlEyeSecretClub}. To obstruct cheat developers from using such memory areas, several checks are put in place by BattlEye through memory scans: anomaly scans, pattern scans and module-specific scans. 

\begin{description}
    \item[Anomaly scans] are used to detect areas of memory with address offsets that are not associated with any known loaded image, meaning all injected libraries or other tampering attempts with the memory of the game. In case unknown memory areas are detected, BattlEye sends a report with information concerning the violated memory areas. 
    \item[Pattern scans] use basic antivirus tactics, comparing loaded modules with known signatures of cheats or methods of injection. Some of these signatures are hardcoded into the BEClient, but most are streamed from the BEServer to the BEClient. BattlEye iterates over the whole game memory and the flagged pages, comparing the streamed signatures with the identified ones. If a known pattern is detected, a report is issued.  
    \item[Module-specific scans] check the memory for any modules that can be used to tamper with the memory of the game. These modules contain checks for specific DLLs, including \mintinline{text}|mmres.dll| and \mintinline{text}|mshtml.dll|. 
    Other specific module identifiers, based on in-memory offsets and timestamps, are streamed from the BEServer during the execution of the game, which are most likely used to detect other libraries used for cheat injection~\cite{web:BattlEyeSecretClub}.
    Additional scans are conducted to inspect other loaded modules, particularly kernel modules. These modules are enumerated by invoking the function \mintinline{c}|NtQuerySystemInformation()|, which provides comprehensive details about all the modules loaded into the system. If a module found in the blocklist is detected within the kernel, such as a known vulnerable driver that has been exploited to inject malicious code into the kernel, BattlEye takes immediate action, terminating the game and transmitting a report to the BEServer \cite{web:GHBattleEye}.
\end{description}

\paragraph{Page Guard}
BattlEye safeguards the memory areas of the game by inspecting memory pages for the presence of the \mintinline{text}|PAGE_GUARD| flag. This flag is a memory protection option that can be enabled when allocating or protecting a page in memory. Once set, it designates a guard page that triggers a \mintinline{text}|STATUS_GUARD_PAGE| exception when accessed, alerting BattlEye to potential unauthorised or malicious activity targeting the protected memory region. Utilising the \mintinline{text}|PAGE_GUARD| flag is beneficial in detecting memory corruption problems or identifying malicious code attempting to write into protected memory regions.\footnote{\phantom{~}\href{https://learn.microsoft.com/en-us/windows/win32/memory/memory-protection-constants}{https://learn.microsoft.com/en-us/windows/win32/memory/memory-protection-constants}} However, it appears that BattlEye does not rely on this feature in practice, only checking if the protection of the requested memory page was changed~\cite{web:BattlEyeSecretClub}.

\paragraph{Window Enumeration}
To detect hooking and, in turn, spoofing of functions in the BattlEye modules, all visible windows on the desktop are enumerated. Overlapping windows are iterated, starting with the highest z-index (corresponding to the uppermost window) and decrementing the index for each window, getting the window handle according to the current one. 
In case fewer than two windows are detected by this method, the anti-cheat assumes its functions were hooked and returned fabricated results, leading to a report~\cite{web:BattlEyeSecretClub}.

\paragraph{Process Enumeration}
On each occasion this module is invoked, the anti-cheat system snapshots all currently running processes, checking each for various anomalies. 
Based on a blocklist of applications commonly used to inject cheats, when such a process is found, BattlEye will report the user~\cite{web:BattlEyeSecretClub}. Apart from scanning for process names, BattlEye also verifies the path of the running process. 


\paragraph{LSASS Checks}\label{subsubsec:BattlEyeLSASSCheck}
The Local Security Authority Subsystem Service (LSASS) validates credentials when a user attempts to log in on a Microsoft Windows OS, enforcing the local security policy and issuing security tokens.\footnote{\phantom{~}\href{https://learn.microsoft.com/en-us/windows-server/security/windows-authentication/credentials-processes-in-windows-authentication}{https://learn.microsoft.com/en-us/windows-server/security/windows-authentication/credentials-processes-in-windows-authentication}} 
Historically, the Local Security Authority (LSA) has been targeted for infecting systems with viruses, taking advantage of its vulnerabilities. 
Due to the responsibilities of the LSA, BattlEye not only checks but hooks the syscalls issued from and to it, redirecting those calls to the BEDaisy driver, which can monitor every process from the kernel and crash the system when an invalid call is issued~\cite{web:BattlEyeSecretClub}.

\paragraph{Networking}
BattlEye conducts system scans to detect all open TCP connections to known cheating sites. These sites are utilised for DRM purposes, which help address the issue of illegal software distribution faced by cheat developers and regular game developers alike. 
When starting the cheat, it must endure such a DRM check by connecting to a server and authenticating the user with valid credentials, ensuring the user paid for the cheats. This connection must often be kept established when the cheat is in use. Open TCP connections are listed in the TCP table of the system and thus are easily readable for the anti-cheat system. A report is generated if an IP address of a known cheating provider is found~\cite{web:BattlEyeSecretClub}.

\paragraph{Tick Scan} \label{subsubsec:BEtickscan}
If this test is invoked, BattlEye will schedule the current thread to sleep for exactly one second. Before the sleep command is issued, it obtains the current (game) tick count and compares it with the tick count after sleeping for the second. If the tick delta is more than 1200 ms, it will generate a report~\cite{web:BattlEyeSecretClub}. This check attempts to detect if an attacker injected code to speed up the game ticks, which could indicate a speed hack, i.\,e.\ the player can move faster in-game than intended. Another way this check can be used is to detect if the game is executed inside a virtual machine, which often has more overhead and thus takes longer to execute the same instructions than bare metal~\cite{web:BattlEyeSecretClubHVDetection}.

\paragraph{Other Methods}
There are other checks this anti-cheat system conducts, mostly scans for applications like 7zip or Visual Studio Code. Additionally, BattlEye regularly transmits information to the server, which includes details about all processes, particularly those with active handles on the game. It also provides game file sizes for integrity checks and reports information regarding hooks detected on any process~\cite{web:BattlEyeSecretClub}. If the BEServer is not certain if a player is cheating, it requests more information by streaming the shellcode to the BEClient, which sends back all information, leading to substantial information exfiltration. 

\subsubsection{BattlEye Self-Protection}\label{subsec:BESelfProtection}
BattlEye not only safeguards the intended game but also protects itself. It extends the established practices outlined in \autoref{subsec:BattlEyeBasics} by employing binary virtualisation and integrity checks. The virtualisation of the images is conducted with \textit{VMProtect}, hardening the reverse-engineering and disassembling process. 
The integrity checks are performed by searching for the file in memory, acquiring a handle and subsequently attempting to open the file. 
This approach allows BattlEye to verify the integrity of the file by confirming its presence in memory and ensuring that it can be successfully accessed. If this operation is successful, BattlEye will get the certificate of the file and compare it against a list of prohibited certificates. In the event of an attempt to load a forbidden file, the anti-cheat system notifies the user regarding the tampering and prohibits the game from loading~\cite{web:BattlEyeSecretClubIntegrityChecks}.

\paragraph{Uninstalling}
Uninstalling the BattlEye anti-cheat system varies for each game, but generally it will uninstall with the game. If not, it can be removed with the script \mintinline{text}|Uninstall_BattlEye.bat|, which is installed along with BattlEye. To remove the service created when installing BattlEye, the command \mintinline{text}|sc delete BEService| must be executed from an elevated command prompt.\footnote{\ \href{https://www.battleye.com/support/faq/}{https://www.battleye.com/support/faq/}} 

\subsection{Easy Anti-Cheat}
EAC is the anti-cheat solution developed by Epic Games, obtained by the company in 2018. It is provided free to use for game developers.\footnote{\ \href{https://www.easy.ac/licensing/}{https://www.easy.ac/licensing/}} 
Well-known games using EAC, amongst others, are \textit{Apex Legends}, \textit{Dead by Daylight} or \textit{Fortnite}. When assessing the security features of EAC, it is considerably more challenging to obtain the reverse-engineered source code compared to BattlEye, due to the increased security measures in the software. 

EAC is designed to implement global bans for players who engage in misconduct by banning their corresponding HWID. This prevents cheaters from creating new accounts and accessing the game from the same machine. In contrast to BattlEye, global bans through EAC are restricted to the game the player has cheated in, not extending to other games using EAC. The HWID used to identify the systems of the players is generated from the following identifiers: 
\begin{description}
    \item[Registry Keys.] Keys containing system hardware identifiers, information about the BIOS, the installed graphics card, and others~\cite{web:ghEACBypass}.
    \item[MAC Address.]\label{item:EACMACAddress} EAC scans the system for all network interface cards (NICs) using their MAC address in the HWID.\footnote{\ \href{https://github.com/ch4ncellor/EAC-Reversal/blob/main/hwid.cpp}{https://github.com/ch4ncellor/EAC-Reversal/blob/main/hwid.cpp}\label{fn:HWID}} 
    \item[Disk Serial Numbers.]\label{item:EACDiskSerials} Every single disk possesses a unique serial number to identify it. EAC incorporates the unique serial numbers of all installed disks in the system as part of the HWID.\footnotemark[\value{footnote}]
    \item[WMI Queries.] The Windows Management Instrumentation (WMI) is used by EAC to query specific identifiers from the actual hardware via the \mintinline{text}|Win32_BaseBoard| class. EAC utilises identifiers from the motherboard and the GPU.\footnotemark[\value{footnote}]
    \item[Processor Features.] CPUs do not have unique identifiers, just the model number and the given feature set. 
    These metrics are included in the generation of the HWID.\footnotemark[\value{footnote}]
\end{description}

While the HWID generation incorporates the metrics mentioned earlier (and possibly others), it is still technically feasible to spoof all of these values. However, this requires significant effort and expertise to do successfully. This system is hence not flawless, but given the dedication it takes to fabricate these values and the secrecy about them from the official publishers, it is deemed good enough to issue effective global bans.

\subsubsection{EAC Anti-Cheat Measures}
With an understanding of the generation of the HWID used for banning players with an unfair advantage, it is possible to delve into the mechanisms EAC employs to detect cheating players. EAC consists of several parts cooperating to protect the game against cheaters. Each time a game protected by EAC is launched, it initiates the download of an up-to-date package that includes the driver module and the game-specific module from the EAC content delivery network. This ensures that the necessary components for EAC's operation, including the latest driver and game module versions, are obtained before the game is executed. Meanwhile, the EAC executable is launched, opening a shared buffer that allows communication between the kernel driver and the user-mode application. The kernel driver is used to monitor the system for various malicious changes. Furthermore, the \mintinline{text}|EasyAntiCheat.dll| is injected into the game, enabling EAC to inspect all changes and input registered in the game~\cite{web:EAC-structure}. The methods employed by EAC to detect cheaters are the following:

\paragraph{Hook Detection and Blocking}
The kernel driver of EAC can detect various hooking techniques, alerting the anti-cheat system when such a hook is detected outside the memory of the game. If a malicious hook is identified, the kernel driver will automatically block this hook and communicate via the shared buffer to generate a report. The purpose of implementing these measures is to impede cheaters from injecting malicious code into the game and to prevent the creation of process handles that could be exploited by external tools to interact with the game~\cite{web:ghEACBypass}.

\paragraph{Memory Scans}\label{subsec:EACMemoryScans}
Like BattlEye, EAC scans the memory of the game, including adjacent memory areas, to detect loaded cheats. The techniques used are very similar to those BattlEye employs~\cite{web:ghEACBypass}. 

\paragraph{Driver Scanning and Logging}
The EAC kernel driver performs scans on the system to identify and block any loaded drivers that may be considered malicious. Specific known drivers it scans for are \mintinline{text}|Dbgv.sys| and \mintinline{text}|PROCMON23.sys|, two drivers of the Sysinternals Suite, as well as \mintinline{text}|dbk64.sys|, the driver for the tool Cheat Engine.
In addition to scanning and blocking suspicious drivers, EAC generates detailed logs that capture information about loaded drivers and loading attempts within the system. These logs are forwarded from the kernel driver to the service executed in user mode, encrypted, and then sent to the EAC servers.

\paragraph{Stack Walking}
EAC applies stack walking to enhance hook detection, a technique to trace the origin of a method call in memory. 
Using the two functions \mintinline{c}|RtlLookupFunctionEntry()| and \mintinline{c}|RtlVirtualUnwind()|, it is possible to obtain the base address of the image hooking the application~\cite{web:ghEACBypass}.

\paragraph{Hypervisor Detection}
EAC uses the feature set of CPUs to detect if it is executed in a virtualised environment or on an OS installed on bare metal, accomplished by executing a single \mintinline{asm}|vmread| instruction upon starting. This instruction attempts to read a field from the virtual machine control structure~\cite[p.\ 2491]{manual:IntelArchitecture}. By verifying the successful execution of the \mintinline{asm}|vmread| instruction without triggering an exception, EAC can determine if it is running inside a virtual machine. If the instruction executes without issues, EAC blocks the game from launching~\cite{web:SecretClubHVDetection}. In addition to the VM test, EAC employs the same tactic as BattlEye, sleeping for one second and measuring the game ticks happening in between, as already described in \autoref{subsubsec:BEtickscan}~\cite{web:SecretClubHVDetection}.

\paragraph{Window Enumeration}
Like BattlEye (see \autoref{subsec:BattlEyeBasics}), EAC employs a technique to enumerate all visible windows during runtime, obtaining the handle and the window name. If a window is detected containing a name corresponding to any cheat, a report is generated~\cite{web:ghEACBypass}.

\paragraph{Manual Mapping Detection}
Manual mapping describes loading a DLL into memory using custom tools instead of Windows API functions. This technique offers the benefit of total control over the location where the library is loaded, circumventing automatic detections from both the OS and the applications involved~\cite{web:GHManualMapping}. 
EAC employs a combined approach of memory scanning to identify manually mapped modules or drivers. The routine begins with scanning processes and threads, during which the associated image bases are saved. Subsequently, the captured base addresses in the virtual memory space are examined for anomalies, such as Portable Executable headers indicating the presence of injected applications within the memory area. Furthermore, like BattlEye, the system verifies adjacent memory regions with the execute bit set, actively searching for any non-module code present~\cite{web:ghEACBypass}. If any modules are detected in the areas outside of the game not associated with a module, the anti-cheat system will dump all strings and syscall invocations found in the section and send them to the developers in an encrypted form.

\paragraph{Handles and Threads}
EAC obtains a list of all open handles on the system, using them to search for any process accessing the game without authorisation. If such a process can be located, the user is flagged. Furthermore, handles pointing to sections in the physical memory are probed to check if any process writes code directly into memory which could be executed in the context of the game or the anti-cheat system~\cite{web:ghEACBypass}. Additionally, EAC scans for all threads running in the kernel or user mode and compares them with defined detections for suspicious threads. If those threads are called from an executable memory region not associated with any loaded module of the EAC system, it generates a report~\cite{web:ghEACBypass}.

\paragraph{Instrumentation Callbacks}
Instrumentation callbacks enable simple syscall monitors from user mode in EAC. Whenever a syscall is executed and the kernel returns to user mode, the anti-cheat system checks for the address of the \mintinline{cpp}|InstrumentationCallback|, which is stored within the \mintinline{cpp}|KPROCESS| structure. If this address points to a valid memory space, the kernel swaps the return address of the function for the one inside the \mintinline{cpp}|InstrumentationCallback| member. The substitution of the return address during the transition from kernel to user mode in EAC occurs only when the \mintinline{asm}|dr7| register is set, triggering a detour. Without setting this register, the return address remains unchanged, and the transition from kernel to user mode proceeds normally without any modifications. When a user executes a game with EAC enabled, the kernel driver creates an instrumentation callback routine that monitors the system calls and other events made from the game.\footnote{\phantom{~}\href{https://www.unknowncheats.me/forum/anti-cheat-bypass/561479-eacs-instrumentation-callback-bypass.html}{https://www.unknowncheats.me/forum/anti-cheat-bypass/561479-eacs-instrumentation-callback-bypass.html}} In this callback function, EAC can scan for any cheating attempts by using techniques like stack walking to find the module that executed the syscall in the first place. 

\paragraph{Game Integrity Checks}
During runtime, EAC employs integrity validation by downloading a catalogue of valid signatures encompassing all game files. This catalogue is used to verify the integrity of game files whenever the game accesses them. Furthermore, the configuration containing the definitions of files to probe is also signed and checked at runtime so that a malicious actor cannot tamper with it. If files experience prohibited modifications mismatching these signatures, EAC restricts the user from playing online games, but allows offline gaming.\footnote{\ \href{https://dev.epicgames.com/docs/game-services/anti-cheat/using-anti-cheat}{https://dev.epicgames.com/docs/game-services/anti-cheat/using-anti-cheat}} 
In addition to its existing detection mechanisms, the EAC driver incorporates checks that prevent the simultaneous execution of games protected by EAC and those protected by BattlEye. Specifically, if the BattlEye kernel driver BEDaisy.sys (as described in \autoref{sec:BattlEye}) is detected to be running on the system, the game protected by EAC will not start~\cite{web:GHEACByPassComment1}. 

\subsubsection{Easy Anti-Cheat Self-Protection}
Like BattlEye, EAC employs self-protecting techniques as well. There is little public knowledge about this topic, but it is generally accepted that these measures are similar to those of BattlEye. EAC also utilises a proprietary virtual machine obfuscator, which conceals the binaries and executes them on non-standard architecture. In order to perform self-checks, EAC employs the use of the virtual \mintinline{asm}|vmcall| instruction. This instruction allows EAC to compare the loaded driver in memory with a clean copy that was read from the disk before the memory mapping process. If the anti-cheat system detects any discrepancies in the integrity of the code, it deliberately crashes the protected application by attempting to access the invalid memory area \mintinline{text}|0xEAC|~\cite{web:EACIntegrityProtection, web:EACIntegrityCheck}.

\paragraph{Uninstalling}
Uninstalling Easy Anti-Cheat from the system happens automatically when the last game using this anti-cheat solution is uninstalled. If remnants remain, EAC can be manually uninstalled by running the installer and selecting the uninstall option. This rids the system of all residues of EAC.\footnote{\ \href{https://www.easy.ac/en-us/support/game/guides/installer/}{https://www.easy.ac/en-us/support/game/guides/installer/}} 

\subsection{FACEIT Anti-Cheat}
FACEIT Anti-Cheat is a third-party service for the games \textit{Counter-Strike: Global Offensive}, \textit{Dota 2} and \textit{Team Fortress 2}, amongst others. It offers ranked leagues with better anti-cheat capabilities than the built-in matchmaking systems in the mentioned games. FACEIT Anti-Cheat uses a combined approach to detect cheating players by collecting data from the locally installed FACEIT Anti-Cheat client and behavioural analysis of players recorded on the server. The latter system is based on the FACEIT Anti-Cheat servers and thus will not be analysed in this article. This approach allows for comparing multiple data sources, making it harder for players to cheat in such matches. 

FACEIT Anti-Cheat heavily restricts the usage of the anti-cheat client with regard to the platform the game is played on and bans players for attempts to circumvent these measures, considering them cheating. The terms of service of FACEIT Anti-Cheat explicitly prohibit the installation of the client, including the anti-cheat system, on a virtual machine. Engaging in such behaviour is considered a violation resulting in a ban from the platform. The attempt to debug or decompile any part of the anti-cheat solution is also considered cheating.\footnote{\ \href{https://support.faceit.com/hc/en-us/articles/360015788779-What-is-deemed-to-be-a-cheat}{https://support.faceit.com/hc/en-us/articles/360015788779-What-is-deemed-to-be-a-cheat}} 
FACEIT Anti-Cheat put these measures in place to protect the anti-cheat from malicious individuals attempting to detect exploits. 

\subsubsection{FACEIT Anti-Cheat Measures}
Consisting of a kernel driver (\mintinline{text}|FACEIT.sys|), the FACEIT Anti-Cheat client and the server-side evaluation, FACEIT Anti-Cheat offers a sophisticated framework to protect the offered games against cheaters. The kernel driver utilises similar anti-cheat tactics as previously discussed anti-cheat systems.\footnote{\ \href{https://support.faceit.com/hc/en-us/articles/9394666828188-What-is-FACEIT-Anti-cheat-and-how-does-it-work}{https://support.faceit.com/hc/en-us/articles/9394666828188-What-is-FACEIT-Anti-cheat-and-how-does-it-work}} The FACEIT Anti-Cheat driver boots with the operating system to ensure the system is in a trusted state. The FACEIT Anti-Cheat kernel driver is designed to constantly collect data about the system, including other running drivers and executed programs, transmitting it to the server when the client starts to detect any illegal behaviour indicators. Known FACEIT Anti-Cheat tactics on the client and the server side are as follows:

\paragraph{Instrumentation Callbacks}
FACEIT Anti-Cheat places instrumentation callbacks on several Windows kernel functions. The anti-cheat system blocks the loading of external modules and DLLs, logs drivers and general loading events via a callback on the \mintinline{c}|LoadImage()| function. Further callbacks are placed on the \mintinline{c}|CreateProcess()| and \mintinline{c}|CreateThread()| functions, including all variantions, which notify the anti-cheat system about created processes or threads for logging and reporting~\cite{web:guidedHackingFACEITAC}.

\paragraph{Driver Detection and Stripping}
The anti-cheat system has built-in detection and unloading capabilities for vulnerable drivers that could inject malicious code for cheats (similar to Vanguard, see \autoref{subsec:VanguardLimitations}). Additionally, mechanisms are in place to locate and close open handles to the protected game to ensure no other application is interfering~\cite{web:guidedHackingFACEITAC}.

\paragraph{Virtualisation Detection}
Due to the strict approach against installation on virtual machines, a detection mechanism against this is in place. FACEIT Anti-Cheat blocks execution with the Windows built-in virtualisation solution Hyper-V enabled, requiring the players to execute custom commands to fully disable it. 
According to FACEIT, it is impossible to “maintain a fair playing environment” with this feature enabled.\footnote{\ \href{https://support.faceit.com/hc/en-us/articles/360019809319-You-need-to-disable-Hyper-V-to-launch-FACEIT-AC}{https://support.faceit.com/hc/en-us/articles/360019809319-You-need-to-disable-Hyper-V-to-launch-FACEIT-AC}} 
They go so far as to instruct players to disable the memory integrity feature of the Windows kernel (which considerably weakens the system by dropping the special protection for high-security processes in memory) so they can better monitor processes like LSASS and CSRSS.

\paragraph{Virtualised Binaries}
FACEIT Anti-Cheat and the client utilise \textit{VMProtect} for binary virtualisation and commercial packers for further obfuscation, similar to all the previous anti-cheat systems~\cite{web:GHFACEITVMProtect}.

%

\paragraph{System Security}
The FACEIT Anti-Cheat system requires the active use of a Trusted Platform Module to store keys to encrypt the data sent to the servers.\footnote{\ \href{https://support.faceit.com/hc/en-us/articles/4407006362002-Error-Please-enable-TPM-2-0-to-continue}{https://support.faceit.com/hc/en-us/articles/4407006362002-Error-Please-enable-TPM-2-0-to-continue}} 
Furthermore, it is necessary to enable the Data Execution Prevention feature to enhance protection in memory, so pages cannot be written to and executed at the same time, blocking tactics to inject cheats into the memory.\footnote{\ \href{https://support.faceit.com/hc/en-us/articles/360017192040-You-need-to-enable-DEP-to-launch-FACEIT-AC}{https://support.faceit.com/hc/en-us/articles/360017192040-You-need-to-enable-DEP-to-launch-FACEIT-AC}} 
In addition to those requirements, FACEIT Anti-Cheat needs the system to be booted with Secure Boot ensuring the integrity of the system beginning with the UEFI. This often requires changes to the UEFI and reformatting of system disks initialised with \mintinline{text}|MBR| instead of \mintinline{text}|GPT|.\footnote{\href{https://support.faceit.com/hc/en-us/articles/4406281700370-Secure-Boot-needs-to-be-enabled-to-launch-FACEIT-AC}{https://support.faceit.com/hc/en-us/articles/4406281700370-Secure-Boot-needs-to-be-enabled-to-launch-FACEIT-AC}} 

Due to FACEIT Anti-Cheat being very secretive about the anti-cheat system and discouraging system analysis, very little data is available. Due to this, it can be assumed that there are additional detection and monitoring features included in the anti-cheat system which are not covered above. 
If players are detected to be cheating, the system flags them for the future, but does not issue instant bans to avoid letting cheaters know which specific action triggered the system. Such flagged accounts are typically banned within one week after the incident. 

To obstruct cheaters from creating new accounts, FACEIT Anti-Cheat uses HWID bans. However, it is worth noting that the HWID generation method used by FACEIT Anti-Cheat is considered weaker than other anti-cheat solutions like Vanguard, as discussed in \autoref{subsec:VanguardHWIDGeneration}. Only two known values are included in the HWID of FACEIT Anti-Cheat: The MAC addresses of the local NICs and the serial number of the first disk in the system. Using these relatively simple indicators, the FACEIT Anti-Cheat kernel driver continuously monitors any changes and spoofing attempts made to those values and notifies the server when the client is executed.

\paragraph{Uninstalling}
FACEIT Anti-Cheat offers no clear instructions for removing the client or the anti-cheat system. Both can be removed via the default Windows uninstalling menu, but there is no guarantee that no traces of the applications will remain on the system. Furthermore, no information is available to revert the changes needed to start the anti-cheat system in the first place, making the whole process very intransparent.

\begin{table*}[bt]
\centering
\scriptsize
\begin{tabular}{r c c c c c c c c}
    \textbf{System} &
    \textbf{Evasion} &
    \textbf{Virtualisation} &
    \textbf{Time of Execution} &
    \textbf{Remote Access} &
    \textbf{Information Exfiltration} &
    \textbf{Network Manipulation} &
    \textbf{Removeability} &
    \textbf{Sum} \\ \toprule
BattlEye          & \xmark & \cmark & \xmark & \cmark & \cmark & \xmark & \xmark & $3$ \\
Easy Anti-Cheat   & \xmark & \cmark & \xmark & \xmark & \cmark & \xmark & \xmark & $2$ \\
FACEIT Anti-Cheat & \cmark & \cmark & \cmark & \xmark & \cmark & \xmark & \cmark & $5$ \\
Vanguard          & \cmark & \cmark & \cmark & \xmark & \cmark & \xmark & \xmark & $4$ \\ 
Flame             & \cmark & \xmark & \xmark & \cmark & \cmark & \cmark & \cmark & $5$ \\
\end{tabular}%
\caption{Overview of the analysed anti-cheat systems, including a comparison with the Flame rootkit.}
\label{tab:AllEvaluation}
\end{table*}

\subsection{Vanguard}
Vanguard, developed by Riot Games specifically for the game \textit{Valorant}, is widely recognised as one of the most intrusive anti-cheat systems. \footnote{\ \href{https://www.gamechampions.com/en/blog/valorant-anti-cheat-vanguard/}{https://www.gamechampions.com/en/blog/valorant-anti-cheat-vanguard/}} It operates upon system boot, ensuring the system starts in a trusted state. Vanguard comprises three distinct components: 
%
%
\begin{description}
    \item[\mintinline{text}|vgk.sys|] is the kernel driver of the anti-cheat system, creating a protected memory area for the user mode part of Vanguard and the game itself. It features user-mode and self-integrity checks and communicates with the user-mode part of Vanguard, enabling command execution from there, including remote update and uninstall methods~\cite{web:vanguard1}.
    \item[\mintinline{text}|vgc.exe|] is the user-mode counterpart to the kernel driver, issuing the commands to execute various scans. It ensures the integrity of the game and the reporting functionality to the game servers. Direct communication with \textit{Valorant} is also implemented~\cite{web:vanguard1}.
    \item[\mintinline{text}|Valorant.exe|] is the game protected by Vanguard. It is executed in the protected memory area created by the kernel driver and continuously checked by the user-mode anti-cheat part of the system~\cite{web:vanguard1}.
\end{description}

Like the previous anti-cheat solutions, Vanguard utilises the HWID for player identification and banning. Riot Games has taken significant measures to protect the details and inner workings of Vanguard. Publicly available information about the anti-cheat system is limited due to the bug bounty program the company offers. 
While the specific details of Vanguard's HWID generation process are not publicly disclosed, certain values are commonly assumed to be used in the generation of the HWID. These values, although not confirmed by Riot Games, are often speculated to include:
\begin{description} \label{subsec:VanguardHWIDGeneration}
    \item[Disk Serials.] Similar to EAC (see \autoref{item:EACDiskSerials}), it is believed that Vanguard incorporates the serial numbers and volume identifiers of the disks present in the system as part of its HWID generation process \cite{web:vanguardHWIDGeneration}.
    \item[Registry Keys.] In case the registry key \mintinline{text}|HKLM\SOFTWARE\| \linebreak \mintinline{text}|Microsoft\Windows NT\CurrentVersion\BuildGUIDEx| is available, it is speculated that Vanguard includes the value of this key in the HWID generation process. The BuildGUIDEx registry key typically contains a unique identifier associated with the specific build of the Windows OS installed~\cite{web:vanguardHWIDGeneration}.
    \item[NVRAM Variables.] Vanguard reportedly incorporates specific information about the platform the system is running on in the HWID generation process. According to~\cite{web:vanguardHWIDGeneration}, this information is sourced from the PlatformData variable, which is said to include a collection of variables set by various devices installed in the system. However, it is important to note that the PlatformData variable itself is not defined in the official UEFI specification \cite[p.\ 1739]{book:UEFISpecification}.
    \item[MAC Addresses.] Like EAC, Vanguard is believed to use the MAC addresses of local NICs in the HWID. 
    \item[Mainboard UUID.] It is also considered plausible that Vanguard leverages the unique identifier stored in the BIOS or UEFI NVRAM of each mainboard during the HWID generation process~\cite{web:vanguardHWIDGeneration}. 
\end{description}

\textit{Valorant} is free-to-play, thus in principle easily allowing cheaters to create new accounts once the previous ones were banned for violations. This emphasises the need for HWID solutions because it is the sole indicator that can identify a system that was cheated on with any certainty.

\subsubsection{Vanguard Anti-Cheat Measures}
Vanguard employs several anti-cheat measures already discussed in detail for EAC, such as instrumentation callbacks or various types of memory scans. To avoid needless repetition, we will focus on measures specific to Vanguard in this subsection.

\paragraph{Shadow Memory}
Vanguard uses a sophisticated method to hide memory pages from other processes and threads executed on the system, protecting them from unauthorised access. This is implemented by hooking the \mintinline{c}|SwapContext()| ioctl function of the Windows kernel, called every time a context switch occurs. By hooking this function, denying access to threads not exclusively allowed is possible. Furthermore, Vanguard uses custom memory addresses in the code based on the shadow memory regions. A crash would occur if any other application attempted to access such a memory address except Vanguard, due to catching occurring page-faults within. 
Resulting from this refined approach, Vanguard is well protected against this kind of cheats by making it very difficult to manipulate the memory of the game~\cite{web:VanguardShadowMemory}.

\paragraph{Virtualisation}
Vanguard and the associated binaries are virtualised and compressed with the proprietary software packer \textit{Packman}, developed by Riot Games.
The packer is utilised to obfuscate and encrypt the data. The essential functions of Vanguard are decrypted at launch, meaning that not all functions are loaded into memory simultaneously. This approach impedes attackers and security researchers from easily understanding and analysing the code~\cite{web:RiotGamesACApproach}.


\paragraph{Remote Controllability}
Vanguard includes features that allow developers to remotely disable and uninstall the anti-cheat system in the event of discovering severe vulnerabilities~\cite{web:vanguard1}. If the feature that enables remote disabling and uninstallation of Vanguard were to be hijacked or exploited by attackers, it poses a severe security risk, granting attackers access to the kernel.

\paragraph{Limitations}\label{subsec:VanguardLimitations}
To enhance the effectiveness of Vanguard, the game is designed not to start if the kernel driver of the anti-cheat system has not been loaded during the boot process. This approach establishes a secure and controlled environment where the anti-cheat system can operate without interference or circumvention attempts. By ensuring that the kernel driver is loaded before the game starts, \textit{Valorant} can prevent other drivers or unauthorised software from being loaded beforehand. If any of those drivers must be loaded when the system is started in the trusted mode, it can be turned off from user mode. This results in the system falling into an untrusted state, which requires a reboot to re-enable the anti-cheat driver. \textit{Valorant} can only be started when the system is in a trusted state.

\paragraph{Uninstalling}
Uninstalling Vanguard is straightforward -- searching for it in the installed applications and selecting uninstall. This removes the files of the anti-cheat system; a reboot is required to unload it from memory, as well~\cite{web:uninstallingVanguard}.

\subsection{Discussion}
We now discuss the findings based on our analysis in \autoref{sec:results}. While no anti-cheat system shows signs of network manipulation, we discovered evidence for all other metrics in at least one of the systems analysed. A summary of the evaluation for the different anti-cheat systems against our metrics is shown in \autoref{tab:AllEvaluation}, including a comparison with \textit{Flame}~\cite{flamerootkit:1}, a rootkit detected by Kaspersky Lab in the Middle East in 2012.

In the category of evasion, Vanguard and FACEIT Anti-Cheat indicated significant traces. Vanguard employs some very intrusive tactics, including evasion tactics circumventing other threads and processes by implementing a custom memory management system, hiding from processes and threads, while FACEIT Anti-Cheat employs vigorous checks for virtualised environments, which is a tactic used by malware and rootkits alike to protect themselves against analysis. While this behaviour was introduced to stop cheaters from tampering with the virtual memory of the game, banning players for merely attempting to play in a VM could be considered unreasonable. Furthermore, the constraint to disable Hyper-V (blocking mundane programs like \textit{WSL-2}) to ensure no virtualisation can occur at all strengthens the evasion tactics. These combined approaches each indicate rootkit-like behaviour.

All of the analysed anti-cheat systems tested positive in the category of virtualisation. All systems employ some kind of virtualisation, each in different ways. BattlEye uses \textit{VMProtect} to virtualise the binaries. Similarly, EAC uses custom software to virtualise the binaries as described in \autoref{subsec:BESelfProtection} for self-protection. FACEIT Anti-Cheat employs heavy obfuscation tactics, as well. Lastly, Vanguard utilises a proprietary packer, including binary encryption by design, which hardens Vanguard against static and dynamic analysis. These methods make it challenging to reverse-engineer and assess them for any security concerns.

Similar to the category of evasion, FACEIT Anti-Cheat and Vanguard are the only two systems designed to execute when the system boots, leading to a positive result for the time of execution metric. FACEIT Anti-Cheat and Vanguard both require the kernel driver to boot with the operating system to check for any other vulnerable drivers to block, ensuring no cheats can be loaded on the system. Furthermore, FACEIT Anti-Cheat requires Secure Boot to be enabled, so no disallowed drivers or applications can be loaded before the OS starts. This design is the most intrusive of all anti-cheat systems. The other two kernel-level anti-cheat systems only start when the protected game is launched, thus posing no threat in this category. 

Concerning remote access methods, traces could only be located in the binary of BattlEye, due to it sending a large amount of information to the BEServer backend. Although the other solutions show traces of such behaviour (e.\,g.\ Vanguard includes features for the developers to disable or uninstall the system remotely in case severe vulnerabilities are discovered), there is no concrete evidence of real remote access and controllability built into the other anti-cheat solution.

Regarding information exfiltration, all of the kernel-level anti-cheat systems indicated positive results for this metric. BattlEye sends a lot of information about the system and the game to the BEServer backend, while EAC utilises various system identifiers to generate the HWID for identification purposes. Additionally, it logs all drivers and modules in memory and sends this information to the server for analysis and monitoring. This information is continuously transmitted to the server, regardless of whether a player is suspected of cheating. If this data, designed to identify a system uniquely, can be accessed by unauthorised people, tracking or impersonating victims on various game servers is possible. Both FACEIT Anti-Cheat and Vanguard employ reporting methods to send system information, such as the HWID, loaded drivers, modules and crash reports, to the game developers for analysis as well, furthermore heavily relying on analysing logs (including loaded drivers, process and thread events, etc.), which are sent from the client to the server. 

Removing the anti-cheat solutions from the system is elementary for most analysed systems except FACEIT Anti-Cheat. Due to the lack of guides describing its proper removal, there is a chance that remnants of it remain active on the disk after removal attempts, thus possibly acting like a rootkit. Also, no additional guides are available for restoring the system to a state before using the software, leaving it in a possible vulnerable state, especially considering the requirement of disabling the memory integrity feature.

\section{Conclusion}
This article analyses the behaviour and tactics of four well-known kernel-level anti-cheat systems widely used in online gaming and compares them to functionalities and methods used in rootkits. The anti-cheat systems BattlEye and Easy Anti-Cheat showed minor similarities to rootkits, which were insufficient to classify them as such according to our metrics. FACEIT Anti-Cheat and Vanguard, however, were identified as rootkit-like applications due to the utilisation of comparable methods. The findings and insights from this research shed light on the thin line between what distinguishes kernel-level anti-cheat systems from rootkits.

The classification of anti-cheat systems as either rootkit-like or not presents significant challenges. We claim that the two anti-cheat systems we classify as rootkits raise severe concerns regarding their potential for malicious abuse and the associated privacy implications. However, although we do not classify the other two systems as rootkit-like, they nonetheless engage in practices that raise questions regarding their appropriateness and integrity.

While our work provides some first answers, it also highlights the need for further research on this topic to gain a more comprehensive understanding of different anti-cheat systems, their tactics, and how to best reconcile the need for robust anti-cheat solutions with respecting user privacy and system integrity. In particular, this article exclusively focussed on anti-cheat systems available on Microsoft Windows, disregarding other OSes. We thus see the need for future research into anti-cheat systems on other operating systems (e.\,g.\ EAC on Proton for Linux) comparing the applied methods to their Windows counterparts.


\begin{acks}
    We are grateful to Tobias Dam, Patrick Kochberger, Robert Luh, Martin Schmiedecker and the anonymous referees for suggesting numerous improvements to both the content and the presentation of this paper.
\end{acks}

\newpage
\bibliographystyle{ACM-Reference-Format}
\bibliography{literature}


\begin{thebibliography}{38}


\ifx \showCODEN    \undefined \def \showCODEN     #1{\unskip}     \fi
\ifx \showDOI      \undefined \def \showDOI       #1{#1}\fi
\ifx \showISBNx    \undefined \def \showISBNx     #1{\unskip}     \fi
\ifx \showISBNxiii \undefined \def \showISBNxiii  #1{\unskip}     \fi
\ifx \showISSN     \undefined \def \showISSN      #1{\unskip}     \fi
\ifx \showLCCN     \undefined \def \showLCCN      #1{\unskip}     \fi
\ifx \shownote     \undefined \def \shownote      #1{#1}          \fi
\ifx \showarticletitle \undefined \def \showarticletitle #1{#1}   \fi
\ifx \showURL      \undefined \def \showURL       {\relax}        \fi
\providecommand\bibfield[2]{#2}
\providecommand\bibinfo[2]{#2}
\providecommand\natexlab[1]{#1}
\providecommand\showeprint[2][]{arXiv:#2}

\bibitem[Beegle(2007)]%
        {article:rootkits2}
\bibfield{author}{\bibinfo{person}{Lynn~Erla Beegle}.}
  \bibinfo{year}{2007}\natexlab{}.
\newblock \showarticletitle{{Rootkits and Their Effects on Information
  Security}}.
\newblock \bibinfo{journal}{\emph{Information Systems Security}}
  \bibinfo{volume}{16}, \bibinfo{number}{3} (\bibinfo{year}{2007}),
  \bibinfo{pages}{164–176}.
\newblock
\showISSN{1065-898X}
\urldef\tempurl%
\url{https://doi.org/10.1080/10658980701402049}
\showDOI{\tempurl}


\bibitem[Blunden(2013)]%
        {blunden2013rootkit}
\bibfield{author}{\bibinfo{person}{Bill Blunden}.}
  \bibinfo{year}{2013}\natexlab{}.
\newblock \bibinfo{booktitle}{\emph{{Rootkit Arsenal: Escape and Evasion in the
  Dark Corners of the System}}}.
\newblock \bibinfo{publisher}{Jones \& Bartlett Learning},
  \bibinfo{address}{Burlington, MA}.
\newblock
\showISBNx{9781449626365}
\showLCCN{2011045666}
\urldef\tempurl%
\url{https://www.jblearning.com/catalog/productdetails/9781449626365}
\showURL{%
\tempurl}


\bibitem[{"bright", "IDontCode", "irql0"}(2021)]%
        {web:EAC-structure}
\bibfield{author}{\bibinfo{person}{{"bright", "IDontCode", "irql0"}}.}
  \bibinfo{year}{2021}\natexlab{}.
\newblock \bibinfo{title}{{EasyAntiCheat Exploit to Inject Unsigned Code into
  Protected Processes}}.
\newblock \bibinfo{howpublished}{Online}.
\newblock
\urldef\tempurl%
\url{https://blog.back.engineering/10/08/2021/}
\showURL{%
\tempurl}


\bibitem["Broihon"(2018)]%
        {web:GHManualMapping}
\bibfield{author}{\bibinfo{person}{"Broihon"}.}
  \bibinfo{year}{2018}\natexlab{}.
\newblock \bibinfo{title}{{Manual Mapping DLL Injection Tutorial - How To
  Manual Map}}.
\newblock \bibinfo{howpublished}{Online}.
\newblock
\urldef\tempurl%
\url{https://guidedhacking.com/threads/manual-mapping-dll-injection-tutorial-how-to-manual-map.10009/}
\showURL{%
\tempurl}


\bibitem["Daax"(2020)]%
        {web:GHFACEITVMProtect}
\bibfield{author}{\bibinfo{person}{"Daax"}.} \bibinfo{year}{2020}\natexlab{}.
\newblock \bibinfo{title}{{Anticheat Faceit Bypass}}.
\newblock \bibinfo{howpublished}{Online}.
\newblock
\urldef\tempurl%
\url{https://guidedhacking.com/threads/anticheat-faceit-bypass.16113/post-89663?referralcode=ON6pj}
\showURL{%
\tempurl}


\bibitem[{"Daax", "iPower", "ajkhoury", "drew"}(2020)]%
        {web:SecretClubHVDetection}
\bibfield{author}{\bibinfo{person}{{"Daax", "iPower", "ajkhoury", "drew"}}.}
  \bibinfo{year}{2020}\natexlab{}.
\newblock \bibinfo{title}{{How Anti-Cheats Detect System Emulation}}.
\newblock \bibinfo{howpublished}{Online}.
\newblock
\urldef\tempurl%
\url{https://secret.club/2020/04/13/how-anti-cheats-detect-system-emulation.html}
\showURL{%
\tempurl}


\bibitem[Eresheim et~al\mbox{.}(2017)]%
        {Eresheim2017}
\bibfield{author}{\bibinfo{person}{Sebastian Eresheim}, \bibinfo{person}{Robert
  Luh}, {and} \bibinfo{person}{Sebastian Schrittwieser}.}
  \bibinfo{year}{2017}\natexlab{}.
\newblock \showarticletitle{{The Evolution of Process Hiding Techniques in
  Malware – Current Threats and Possible Countermeasures}}.
\newblock \bibinfo{journal}{\emph{Journal of Information Processing}}
  \bibinfo{volume}{25} (\bibinfo{year}{2017}), \bibinfo{pages}{866--874}.
\newblock
\urldef\tempurl%
\url{https://doi.org/10.2197/ipsjjip.25.866}
\showDOI{\tempurl}


\bibitem[Fritsch(2008)]%
        {thesis:rootkits2}
\bibfield{author}{\bibinfo{person}{Hagen Fritsch}.}
  \bibinfo{year}{2008}\natexlab{}.
\newblock \emph{\bibinfo{title}{{Analysis and Detection of Virtualization-Vased
  Rootkits}}}.
\newblock Bachelor's thesis. \bibinfo{school}{{Technical University of
  Munich}}.
\newblock
\urldef\tempurl%
\url{https://www.nm.ifi.lmu.de/pub/Fopras/frit08/PDF-Version/frit08.pdf}
\showURL{%
\tempurl}


\bibitem["h4x0!2"(2023)]%
        {web:vanguardHWIDGeneration}
\bibfield{author}{\bibinfo{person}{"h4x0!2"}.} \bibinfo{year}{2023}\natexlab{}.
\newblock \bibinfo{title}{{Data Vanguard Is Grabbing to HWID Ban}}.
\newblock \bibinfo{howpublished}{Online}.
\newblock
\urldef\tempurl%
\url{https://www.unknowncheats.me/forum/valorant/567650-data-vanguard-grabbing-hwid-ban.html}
\showURL{%
\tempurl}


\bibitem[Intel(2023)]%
        {manual:IntelArchitecture}
\bibfield{author}{\bibinfo{person}{Intel}.} \bibinfo{year}{2023}\natexlab{}.
\newblock \bibinfo{booktitle}{\emph{{Intel® 64 and IA-32 Architectures
  Software Developer’s Manuals}}}.
\newblock Intel.
\newblock
\urldef\tempurl%
\url{https://www.intel.com/content/www/us/en/developer/articles/technical/intel-sdm.html}
\showURL{%
\tempurl}


\bibitem["iPower"(2020)]%
        {web:EACIntegrityCheck}
\bibfield{author}{\bibinfo{person}{"iPower"}.} \bibinfo{year}{2020}\natexlab{}.
\newblock \bibinfo{title}{{CVEAC-2020: Bypassing EasyAntiCheat Integrity
  Checks}}.
\newblock \bibinfo{howpublished}{Online}.
\newblock
\urldef\tempurl%
\url{https://secret.club/2020/04/08/eac_integrity_check_bypass.html}
\showURL{%
\tempurl}


\bibitem[Jiang(2006)]%
        {phd:rootkits1}
\bibfield{author}{\bibinfo{person}{Xuxian Jiang}.}
  \bibinfo{year}{2006}\natexlab{}.
\newblock \emph{\bibinfo{title}{{Enabling Internet Worms and Malware
  Investigation and Defense Using Virtualization}}}.
\newblock PhD thesis. \bibinfo{school}{Purdue University}.
\newblock
\urldef\tempurl%
\url{https://docs.lib.purdue.edu/dissertations/AAI3251634/}
\showURL{%
\tempurl}


\bibitem[Joy et~al\mbox{.}(2011)]%
        {Joy2011RootkitDM}
\bibfield{author}{\bibinfo{person}{Jestin Joy}, \bibinfo{person}{Anita John},
  {and} \bibinfo{person}{James Joy}.} \bibinfo{year}{2011}\natexlab{}.
\newblock \showarticletitle{{Rootkit Detection Mechanism: A Survey}}. In
  \bibinfo{booktitle}{\emph{Proceedings of the First International Conference
  on Parallel Distributed Computing Technologies and Applications}}
  (Tirunelveli) \emph{(\bibinfo{series}{PDCTA 2011/Communications in Computer
  and Information Science, vol.\ 203})}. \bibinfo{publisher}{Springer},
  \bibinfo{address}{Berlin/Heidelberg}, \bibinfo{pages}{366–374}.
\newblock
\urldef\tempurl%
\url{https://doi.org/10.1007/978-3-642-24037-9_36}
\showDOI{\tempurl}


\bibitem[Lehtonen(2020)]%
        {masters:lethonen}
\bibfield{author}{\bibinfo{person}{Samuli Lehtonen}.}
  \bibinfo{year}{2020}\natexlab{}.
\newblock \emph{\bibinfo{title}{{Comparative Study of Anti-Cheat Methods in
  Video Games}}}.
\newblock \bibinfo{thesistype}{Master's\ thesis}. \bibinfo{school}{University
  of Helsinki}.
\newblock
\urldef\tempurl%
\url{https://helda.helsinki.fi/items/b1141406-eb65-48a5-8922-d1b23d4cfe51}
\showURL{%
\tempurl}


\bibitem[Li et~al\mbox{.}(2011)]%
        {Li2011AnOO}
\bibfield{author}{\bibinfo{person}{Xiang Li}, \bibinfo{person}{Yan Wen},
  \bibinfo{person}{Minhuan Huang}, {and} \bibinfo{person}{Qiang Liu}.}
  \bibinfo{year}{2011}\natexlab{}.
\newblock \showarticletitle{{An Overview of Bootkit Attacking Approaches}}. In
  \bibinfo{booktitle}{\emph{Proceedings of the 2011 Seventh International
  Conference on Mobile Ad-Hoc and Sensor Networks}} (Beijing)
  \emph{(\bibinfo{series}{MSN 2011})}. \bibinfo{publisher}{IEEE},
  \bibinfo{address}{New York, NY}, \bibinfo{pages}{428–431}.
\newblock
\urldef\tempurl%
\url{https://doi.org/10.1109/MSN.2011.19}
\showDOI{\tempurl}


\bibitem[Liu et~al\mbox{.}(2012)]%
        {Leian:1}
\bibfield{author}{\bibinfo{person}{Leian Liu}, \bibinfo{person}{Zuanxing Yin},
  \bibinfo{person}{Yuli Shen}, {and} \bibinfo{person}{Haitao Lin}.}
  \bibinfo{year}{2012}\natexlab{}.
\newblock \showarticletitle{Research and Design of Rootkit Detection Method}.
\newblock \bibinfo{journal}{\emph{Physics Procedia}}  \bibinfo{volume}{33}
  (\bibinfo{year}{2012}), \bibinfo{pages}{852–857}.
\newblock
\urldef\tempurl%
\url{https://doi.org/10.1016/j.phpro.2012.05.145}
\showDOI{\tempurl}


\bibitem[Maario et~al\mbox{.}(2011)]%
        {paper:RedefiningKernelLevelACs}
\bibfield{author}{\bibinfo{person}{Anton Maario}, \bibinfo{person}{Vinod
  Shukla}, \bibinfo{person}{A. Ambikapathy}, {and} \bibinfo{person}{Purushottam
  Sharma}.} \bibinfo{year}{2011}\natexlab{}.
\newblock \showarticletitle{{Redefining the Risks of Kernel-Level Anti-Cheat in
  Online Gaming}}. In \bibinfo{booktitle}{\emph{Proceedings of the 2021 8th
  International Conference on Signal Processing and Integrated Networks}}
  (Noida) \emph{(\bibinfo{series}{SPIN 2021})}. \bibinfo{publisher}{IEEE},
  \bibinfo{address}{New York, NY}, \bibinfo{pages}{676–680}.
\newblock
\urldef\tempurl%
\url{https://doi.org/10.1109/SPIN52536.2021.9566108}
\showDOI{\tempurl}


\bibitem[Major(2015)]%
        {major2015taxonomic}
\bibfield{author}{\bibinfo{person}{Maxine Major}.}
  \bibinfo{year}{2015}\natexlab{}.
\newblock \emph{\bibinfo{title}{{A Taxonomic Evaluation of Rootkit Deployment,
  Behavior and Detection}}}.
\newblock \bibinfo{thesistype}{Master's\ thesis}. \bibinfo{school}{{University
  of Idaho}}.
\newblock
\urldef\tempurl%
\url{https://objects.lib.uidaho.edu/etd/pdf/Major_idaho_0089N_10700.pdf}
\showURL{%
\tempurl}


\bibitem[Matrosov et~al\mbox{.}(2019)]%
        {matrosov2019rootkits}
\bibfield{author}{\bibinfo{person}{Alex Matrosov}, \bibinfo{person}{Eugene
  Rodionov}, {and} \bibinfo{person}{Sergey Bratus}.}
  \bibinfo{year}{2019}\natexlab{}.
\newblock \bibinfo{booktitle}{\emph{{Rootkits and Bootkits: Reversing Modern
  Malware and Next Generation Threats}}}.
\newblock \bibinfo{publisher}{No Starch Press}, \bibinfo{address}{San
  Francisco, CA}.
\newblock
\showISBNx{9781593278830}
\showLCCN{2017048113}
\urldef\tempurl%
\url{https://nostarch.com/rootkits}
\showURL{%
\tempurl}


\bibitem[Mysliwietz(2020)]%
        {thesis:RootkitStealthTactics}
\bibfield{author}{\bibinfo{person}{Egidius Mysliwietz}.}
  \bibinfo{year}{2020}\natexlab{}.
\newblock \emph{\bibinfo{title}{{Identifying Rootkit Stealth Strategies}}}.
\newblock Bachelor's thesis. \bibinfo{school}{{Radboud University}}.
\newblock
\urldef\tempurl%
\url{https://www.cs.ru.nl/bachelors-theses/2020/Egidius_Mysliwietz___1000796___Identifying_rootkit_stealth_strategies.pdf}
\showURL{%
\tempurl}


\bibitem[Orland(4 14)]%
        {web:vanguard1}
\bibfield{author}{\bibinfo{person}{Kyle Orland}.}
  \bibinfo{year}{2020-04-14}\natexlab{}.
\newblock \showarticletitle{{Ring 0 of Fire: Does Riot Games’ New Anti-Cheat
  Measure Go Too Far?}}
\newblock \bibinfo{journal}{\emph{Ars Technica}} (\bibinfo{year}{2020-04-14}).
\newblock
\urldef\tempurl%
\url{https://arstechnica.com/gaming/2020/04/ring-0-of-fire-does-riot-games-new-anti-cheat-measure-go-too-far/}
\showURL{%
\tempurl}


\bibitem["Rake"(2015)]%
        {web:GHBattleEye}
\bibfield{author}{\bibinfo{person}{"Rake"}.} \bibinfo{year}{2015}\natexlab{}.
\newblock \bibinfo{title}{{Anticheat Battleye Bypass Overview}}.
\newblock \bibinfo{howpublished}{Online}.
\newblock
\urldef\tempurl%
\url{https://guidedhacking.com/threads/anticheat-battleye-bypass-overview.11602/}
\showURL{%
\tempurl}


\bibitem["Rake"(2018)]%
        {web:guidedHackingFACEITAC}
\bibfield{author}{\bibinfo{person}{"Rake"}.} \bibinfo{year}{2018}\natexlab{}.
\newblock \bibinfo{title}{{Anticheat Faceit Bypass}}.
\newblock \bibinfo{howpublished}{Online}.
\newblock
\urldef\tempurl%
\url{https://guidedhacking.com/threads/anticheat-faceit-bypass.16113/}
\showURL{%
\tempurl}


\bibitem["Rake"(2020)]%
        {web:ghEACBypass}
\bibfield{author}{\bibinfo{person}{"Rake"}.} \bibinfo{year}{2020}\natexlab{}.
\newblock \bibinfo{title}{{How to Bypass EAC - Easy Anti Cheat}}.
\newblock \bibinfo{howpublished}{Online}.
\newblock
\urldef\tempurl%
\url{https://guidedhacking.com/threads/how-to-bypass-eac-easy-anti-cheat.15956/}
\showURL{%
\tempurl}


\bibitem[Rendenbach(2022)]%
        {bakk:Rendenbach}
\bibfield{author}{\bibinfo{person}{Caroline~Andrea Rendenbach}.}
  \bibinfo{year}{2022}\natexlab{}.
\newblock \emph{\bibinfo{title}{{Anti-Cheating Measures in Video Games}}}.
\newblock Bachelor's thesis. \bibinfo{school}{{Technical University of
  Munich}}.
\newblock
\urldef\tempurl%
\url{https://collab.dvb.bayern/download/attachments/77832800/main.pdf}
\showURL{%
\tempurl}


\bibitem[{Riot Games}(2018)]%
        {web:RiotGamesACApproach}
\bibfield{author}{\bibinfo{person}{{Riot Games}}.}
  \bibinfo{year}{2018}\natexlab{}.
\newblock \bibinfo{title}{{Riot's Approach to Anti-Cheat}}.
\newblock \bibinfo{howpublished}{Online}.
\newblock
\urldef\tempurl%
\url{https://technology.riotgames.com/news/riots-approach-anti-cheat}
\showURL{%
\tempurl}


\bibitem[Rolles(2009)]%
        {rolles2009unpacking}
\bibfield{author}{\bibinfo{person}{Rolf Rolles}.}
  \bibinfo{year}{2009}\natexlab{}.
\newblock \showarticletitle{{Unpacking Virtualization Obfuscators}}. In
  \bibinfo{booktitle}{\emph{Proceedings of the 3rd USENIX Workshop on Offensive
  Technologies}} (Montreal) \emph{(\bibinfo{series}{WOOT '09})}.
  \bibinfo{publisher}{USENIX Association}, \bibinfo{address}{Berkeley, CA},
  \bibinfo{pages}{261--266}.
\newblock
\urldef\tempurl%
\url{https://www.usenix.org/legacy/events/woot09/tech/full_papers/rolles.pdf}
\showURL{%
\tempurl}


\bibitem["SaltyPaster"(2021)]%
        {web:GHEACByPassComment1}
\bibfield{author}{\bibinfo{person}{"SaltyPaster"}.}
  \bibinfo{year}{2021}\natexlab{}.
\newblock \bibinfo{title}{{How to Bypass EAC - Easy Anti Cheat}}.
\newblock \bibinfo{howpublished}{Online}.
\newblock
\urldef\tempurl%
\url{https://guidedhacking.com/threads/how-to-bypass-eac-easy-anti-cheat.15956/post-105040?referralcode=ON6pj}
\showURL{%
\tempurl}


\bibitem[Silva(2022)]%
        {masters:silva}
\bibfield{author}{\bibinfo{person}{José~Nuno Silva}.}
  \bibinfo{year}{2022}\natexlab{}.
\newblock \emph{\bibinfo{title}{{Towards Automated Server-side Video Game Cheat
  Detection}}}.
\newblock \bibinfo{thesistype}{Master's\ thesis}. \bibinfo{school}{University
  of Porto}.
\newblock
\urldef\tempurl%
\url{https://repositorio-aberto.up.pt/bitstream/10216/142935/2/572983.pdf}
\showURL{%
\tempurl}


\bibitem["Sinclairq"(2022)]%
        {web:EACIntegrityProtection}
\bibfield{author}{\bibinfo{person}{"Sinclairq"}.}
  \bibinfo{year}{2022}\natexlab{}.
\newblock \bibinfo{title}{{A Bank Vault’s Self-Integrity Circumvented by an
  Underway Passage: How EasyAntiCheat’s Driver Self-Integrity Can Be
  Compromised Through Call Hierarchy}}.
\newblock \bibinfo{howpublished}{Online}.
\newblock
\urldef\tempurl%
\url{https://secret.club/2020/04/08/eac_integrity_check_bypass.html}
\showURL{%
\tempurl}


\bibitem[{UEFI Forum, Inc.}(2019)]%
        {book:UEFISpecification}
\bibfield{author}{\bibinfo{person}{{UEFI Forum, Inc.}}}
  \bibinfo{year}{2019}\natexlab{}.
\newblock \bibinfo{booktitle}{\emph{{Unified Extensible Firmware Interface
  (UEFI) Specification}}}.
\newblock {Unified Extensible Firmware Interface (UEFI) Forum}.
\newblock
\urldef\tempurl%
\url{https://uefi.org/specifications}
\showURL{%
\tempurl}


\bibitem[Virvilis and Gritzalis(2013)]%
        {flamerootkit:1}
\bibfield{author}{\bibinfo{person}{Nikos Virvilis} {and}
  \bibinfo{person}{Dimitris Gritzalis}.} \bibinfo{year}{2013}\natexlab{}.
\newblock \showarticletitle{{The Big Four – What We Did Wrong in Advanced
  Persistent Threat Detection?}}. In \bibinfo{booktitle}{\emph{Proceedings of
  the 2013 International Conference on Availability, Reliability and Security}}
  (Regensburg) \emph{(\bibinfo{series}{ARES 2013})}. \bibinfo{publisher}{IEEE},
  \bibinfo{address}{New York, NY}, \bibinfo{pages}{248–254}.
\newblock
\urldef\tempurl%
\url{https://doi.org/10.1109/ARES.2013.32}
\showDOI{\tempurl}


\bibitem["vmcall"(2019)]%
        {web:BattlEyeSecretClub}
\bibfield{author}{\bibinfo{person}{"vmcall"}.} \bibinfo{year}{2019}\natexlab{}.
\newblock \bibinfo{title}{{BattlEye Anti-Cheat: Analysis and Mitigation}}.
\newblock \bibinfo{howpublished}{Online}.
\newblock
\urldef\tempurl%
\url{https://secret.club/2019/02/10/battleye-anticheat.html}
\showURL{%
\tempurl}


\bibitem["vmcall"(2020)]%
        {web:BattlEyeSecretClubHVDetection}
\bibfield{author}{\bibinfo{person}{"vmcall"}.} \bibinfo{year}{2020}\natexlab{}.
\newblock \bibinfo{title}{{BattlEye Hypervisor Detection}}.
\newblock \bibinfo{howpublished}{Online}.
\newblock
\urldef\tempurl%
\url{https://secret.club/2020/01/12/battleye-hypervisor-detection.html}
\showURL{%
\tempurl}


\bibitem["whatacoolwitch"(2021)]%
        {web:uninstallingVanguard}
\bibfield{author}{\bibinfo{person}{"whatacoolwitch"}.}
  \bibinfo{year}{2021}\natexlab{}.
\newblock \bibinfo{title}{{Uninstalling and Disabling Riot Vanguard}}.
\newblock \bibinfo{howpublished}{Online}.
\newblock
\urldef\tempurl%
\url{https://support-valorant.riotgames.com/hc/en-us/articles/360044648213-Uninstalling-and-Disabling-Riot-Vanguard}
\showURL{%
\tempurl}


\bibitem["whatacoolwitch"(2022)]%
        {web:VanguardOverview}
\bibfield{author}{\bibinfo{person}{"whatacoolwitch"}.}
  \bibinfo{year}{2022}\natexlab{}.
\newblock \bibinfo{title}{{What Is Vanguard?}}
\newblock \bibinfo{howpublished}{Online}.
\newblock
\urldef\tempurl%
\url{https://support-valorant.riotgames.com/hc/en-us/articles/360046160933-What-is-Vanguard-}
\showURL{%
\tempurl}


\bibitem["Xyrem"(2023)]%
        {web:VanguardShadowMemory}
\bibfield{author}{\bibinfo{person}{"Xyrem"}.} \bibinfo{year}{2023}\natexlab{}.
\newblock \bibinfo{title}{{In-Depth Analysis on Valorant’s Guarded Regions}}.
\newblock \bibinfo{howpublished}{Online}.
\newblock
\urldef\tempurl%
\url{https://reversing.info/posts/guardedregions/}
\showURL{%
\tempurl}


\bibitem["yousif"(2020)]%
        {web:BattlEyeSecretClubIntegrityChecks}
\bibfield{author}{\bibinfo{person}{"yousif"}.} \bibinfo{year}{2020}\natexlab{}.
\newblock \bibinfo{title}{{Bypassing BattlEye from User-Mode}}.
\newblock \bibinfo{howpublished}{Online}.
\newblock
\urldef\tempurl%
\url{https://secret.club/2020/02/26/be_umode.html}
\showURL{%
\tempurl}


\end{thebibliography}

\end{document}